# The Bubble regime of laser-plasma acceleration: monoenergetic electrons and the scalability.

A. Pukhov††, S. Gordienko†, S. Kiselev†, and I. Kostyukov‡

†Institute for theoretical physics I
University of Duesseldorf, 40225 Duesseldorf, Germany
‡Institute for applied physics
Nizhni Novgorod, Russia

**Abstract.** The Bubble regime of electron acceleration in ultra-relativistic laser plasma is considered. It has been shown that the bubble can produce ultra-short dense bunches of electrons with quasi-monoenergetic energy spectra. The first experiment in this regime done at LOA has confirmed the peaked electron spectrum (J. Faure, et al., *submitted*, 2004). The generated electron bunch may have density an order of magnitude higher than that of the background plasma. The bubble is able to guide the laser pulse over many Rayleigh lengths, thus no preformed plasma channel is needed for high-energy particle acceleration in the bubble regime. In the present work we discuss a simple analytical model for the bubble fields as well as the scaling laws.

† To whom correspondence should be addressed (pukhov@thphy.uni-duesseldorf.de)

## 1. Introduction

The recent advance in laser technology offers new possibilities for efficient, compact accelerators [1], advanced fusion concepts [2], and new generation of radiation sources [3]. One of the promising schemes is the high-gradient laser wake field acceleration (LWFA) of charged particles in plasmas [5]. When a laser pulse propagates through underdense plasma, it excites a running plasma wave oscillating with the frequency $\omega_p = (4\pi e^2 n_0/\gamma m)^{1/2}$, where $e$, $m$, and $n_0$ denote charge, mass, and density of electrons, respectively; $\gamma$ is the electron relativistic factor. The wave trails the laser pulse with phase velocity set by the laser pulse group velocity $v_{ph}^{wake} = v_0 \simeq c(1 - \omega_p^2/2\omega^2)$, where $\omega$ is the laser frequency. A relativistic electron can ride this plasma wave, staying in-phase with the longitudinal electric field and be accelerated to high energies.

The laser pulse can excite the plasma wave in different ways [4]. The excitation is most efficient when the laser pulse duration is of the order of the plasma wavelength $\lambda_p = 2\pi c/\omega_p$. Taking a plane laser pulse with the normalized intensity profile $a^2 = a_0^2 \cos^2 \pi\zeta/2L$ for $-L < \zeta = z - ct < L$, one finds that the wake field reaches the maximum $E_{\max}/E_0 = a_0^2/(2 + 2a_0^2)$, when the laser pulse full width at half maximum (FWHM) is $L = \lambda_p/2$ [5]. Here $E_0 = mc\omega_p/e$ normalizes the electric field of the plasma wave, and $a = eA_0/mc^2$ is the normalized amplitude of the laser vector potential. The pattern of wake field excitation differs significantly for laser pulses longer and shorter than the plasma period. The long laser pulse gets self-modulated with the plasma period, and the resonance between this self-modulation and the plasma frequency leads to effective wake field excitation. The corresponding regime is called self-modulated laser wake field acceleration (SM-LWFA) [6]. Long laser pulses, however, experience not only the one-dimensional self-modulation, but get self-focused and form relativistic channels in the plasma [7, 8].

Trapping of electrons in the plasma waves is a key issue for LWFA. Injection and acceleration of external beams has been demonstrated experimentally [9]. Creation of trapped electrons inside the wave bucket has been proposed with the application of supplementary laser pulses [10, 11]. The wavebreaking can also lead to self-trapping and acceleration of electrons by the plasma wave. It should be noted, however, that the most experimentally observed spectra of accelerated electrons were exponentially decaying [12, 13].

In the present paper, we focus on laser-plasma interaction in the "bubble" regime recently proposed by A. Pukhov and J. Meyer-ter-Vehn [14]. It has been observed in 3D Particle-in-Cell (PIC) simulations for ultra-relativistically intense laser pulses shorter than $\lambda_p$. These laser pulses are intense enough to break the plasma wave already after the first oscillation. The main features of the bubble regime are the following: (i) a cavity free from cold plasma electrons is formed behind the laser pulse instead of a periodic plasma wave; (ii) a dense bunch of relativistic electrons with a monoenergetic spectrum is self-generated; (iii) the laser pulse propagates many Rayleigh lengths in the homogeneous plasma without a significant diffraction. These features are absent in the ordinary regime of laser wake field acceleration [5].

## 2. NUMERICAL SIMULATIONS

For the simulations, we use the fully electromagnetic 3D PIC code Virtual Laser-Plasma Laboratory [15]. The incident laser pulse is circularly polarized, has the Gaussian envelope $a(t,r) = A_0 \exp(-r_\perp^2/r_L^2 - t^2/T_L^2)$, and the wavelength $\lambda = 0.82$ $\mu$m.

In the first simulation, the laser pulse parameters were $r_L = 10\lambda$, $cT_L = 4\lambda$, $a_0 = 10$. The vacuum Reyleigh length for these parameters is $Z_R \approx 300\lambda$. The pulse propagates in a plasma with the density $n_0 = 6.1 \times 10^{-3} n_c$, where $n_c = (m\omega^2/4\pi e^2)^{1/2}$ is the critical density. The plasma density distribution observed in the simulation is shown in Fig. 1 at two instants of time: (a) when the laser pulse has passed $l_{int}^1 = 25c/\omega_p \simeq 50\lambda$ and (b) $l_{int}^2 = 442c/\omega_p \simeq 900\lambda \simeq 3Z_R$ in plasma. These density distributions are very typical for the bubble regime. It is seen from Fig. 1 that the wake behind the laser pulse takes the form of a solitary cavity, which is free from plasma electrons. The cavity is surrounded by a high density sheath of the compressed electron fluid. At later times, Fig. 1(b), a beam of accelerated electrons grows from the bubble base. Simultaneously, the bubble size increases.

The electron dynamics is defined by the laser ponderomotive force and the electromagnetic fields pertinent to the bubble density patterns. It is seen from Fig. 1 that there are roughly three patterns: (i) the electron plasma cavity with the large ion charge; (ii) the electron sheath around the cavity forming the bubble boundary; (iii) the bunch of accelerated electrons growing behind the laser pulse in the cavity. The density of the electron sheath peaks at the head of the laser pulse and at the base of the cavity. These density peaks are formed by the relativistic electrons with $v \simeq v_0$. The bubble base is the source of electrons, which get trapped and accelerated to $\gamma \gg \gamma_0$, where $\gamma_0 = (1 - v_0^2/c^2)^{1/2}$ is the relativistic gamma-factor of the laser pulse.

## 3. FIELDS INSIDE RELATIVISTIC CAVITY

Here we develop a phenomenological theory of the bubble and approximate the electron cavity by a sphere [17]. Before considering the relativistic cavity moving in plasma we summarize the results for fields within an ionic sphere either at rest, or relativistically moving. The electromagnetic field of the uniformly charged sphere at rest is purely electrostatic. The electric field and the scalar potential inside the sphere with radius $R$ and with the charge density $|e|n_0$ are [18]

$$\mathbf{E} = \frac{\mathbf{r}}{3}, \quad \mathbf{B} = 0, \quad \varphi = 1 + \frac{R^2}{6} - \frac{r^2}{6}, \tag{1}$$

where we choose that the potential is equal to unity at the sphere boundary. We use dimensionless units, normalizing the time to $\omega_p^{-1}$, the lengths to $c/\omega_p$, the velocity to $c$, the electromagnetic fields to $mc\omega_p/|e|$, and the electron density, $n$, to the background density $n_0$.

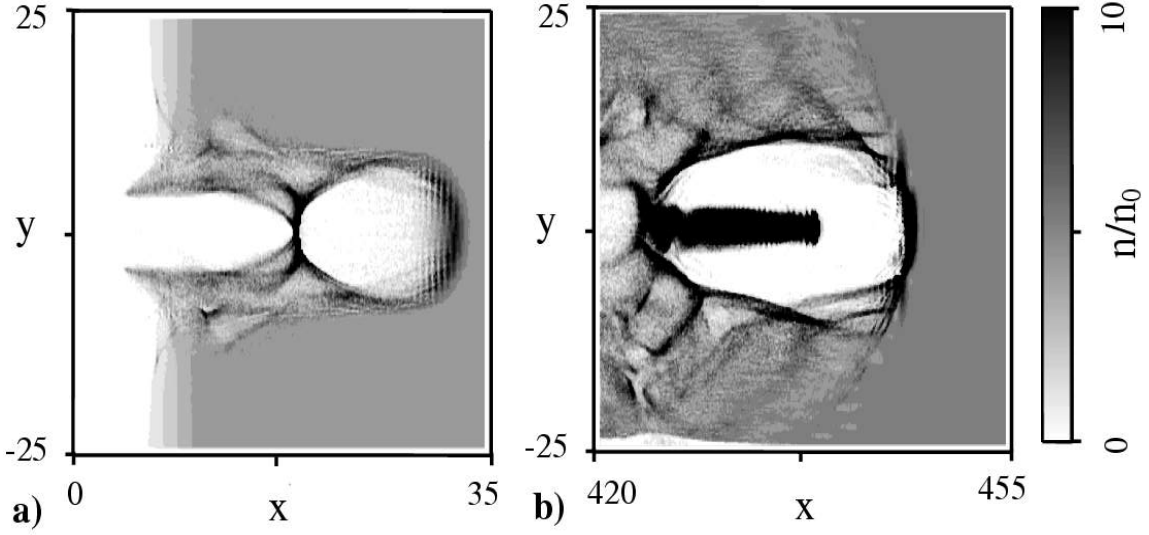

**Figure 1.** On-axis cuts of the electron density $n$ in the $x - y$ plane from the PIC simulation at the times when the laser pulse passed about (a) $L_1 = 25c/\omega_p \simeq 50\lambda$ and (b) $L_2 = 442c/\omega_p \simeq 900\lambda$. The coordinates are given in $c/\omega_p$.

If the ionic sphere runs with the relativistic velocity $v_0 \simeq 1$ along $x$-axis then the fields inside the sphere are

$$E_x = 0, \quad E_y = B_z = \frac{y}{2}, \tag{2}$$

$$B_x = 0, \quad E_z = -B_y = \frac{z}{2}, \tag{3}$$

where the terms, which are proportional to $\gamma_0^{-2} = 1-v_0^2 \ll 1$, are neglected. The Lorentz force on the relativistic electron moving inside the sphere with velocity $v_x = v = -1$ is

$$F_x = 0, \quad F_y = -E_y - B_z = -y, \tag{4}$$

$$F_z = -E_z + B_y = -z, \tag{5}$$

while it is negligible in the limit $v_0 = v_x = 1$ because of relativistic compensation of the electrostatic force by the self-magnetic force [19].

Now we are interested in the question what are the fields inside a spherical electron cavity moving in plasma. This cavity is similar to the hole in semiconductor physics [20]. Contrary to the case discussed above, the ions are now immobile in the cavity while the cavity runs with the relativistic velocity $v_0 \simeq 1$ along $x$-axis. The ion dynamics is neglected because the cavity radius is assumed to be smaller than the ion response length $\simeq c/\omega_{pi}$, where $\omega_{pi} = (4\pi e^2 n_0/M)$ is the ion plasma frequency and $M$ is the ion mass. To calculate the fields we rewrite the Maxwell equations in terms of potentials using the following convenient gauge

$$A_x = -\varphi. \tag{6}$$

We get

$$\triangle \Phi = 1 - n\left(1 - \frac{p_x}{\gamma}\right) + \left(\frac{\partial}{\partial t} + \frac{\partial}{\partial x}\right)(\nabla \cdot \mathbf{A})$$
$$+ \frac{1}{2}\frac{\partial}{\partial t}\left(\frac{\partial}{\partial t} - \frac{\partial}{\partial x}\right)\Phi, \tag{7}$$

$$\nabla \times \nabla \times \mathbf{A} + n\frac{\mathbf{p}}{\gamma} + \frac{\partial}{\partial t}\left(\frac{\partial \mathbf{A}}{\partial t} - \frac{\nabla \Phi}{2}\right) = 0. \tag{8}$$

Here we use the wake field potential $\Phi = A_x - \varphi$ instead of the scalar one, $n$ is the electron density and $\mathbf{p}$ is the electron momentum.

Then we use a quasistatic approximation assuming that all quantities depend on $\zeta = x - v_0 t$ instead of $x$ and $t$. The Maxwell equations reduce to the form

$$\triangle \Phi = \frac{3}{2}(1-n) + n\frac{p_x}{\gamma} - \frac{1}{2}\frac{\partial}{\partial \xi}(\nabla_\perp \cdot \mathbf{A}_\perp) \tag{9}$$

$$\triangle_\perp \mathbf{A}_\perp - \nabla_\perp(\nabla_\perp \cdot \mathbf{A}_\perp) = n\frac{\mathbf{p}_\perp}{\gamma} + \frac{1}{2}\nabla_\perp\frac{\partial \Phi}{\partial \xi}, \tag{10}$$

where the terms proportional to $\gamma_0^{-2} \ll 1$, are neglected. Inside the cavity ($n = 0$) we get

$$\triangle \Phi = \frac{3}{2} - \frac{1}{2}\frac{\partial}{\partial \xi}(\nabla_\perp \cdot \mathbf{A}_\perp), \tag{11}$$

$$\triangle_\perp \mathbf{A}_\perp - \nabla_\perp(\nabla_\perp \cdot \mathbf{A}_\perp) = \frac{1}{2}\nabla_\perp\frac{\partial \Phi}{\partial \xi}. \tag{12}$$

The solution of Eqs. (11) and (12) with spherical symmetry is

$$\Phi = 1 - \frac{R^2}{4} + \frac{r^2}{4}, \quad A_z = -\varphi = \frac{\Phi}{2}, \quad \mathbf{A}_\perp = 0, \tag{13}$$

where $R$ is the radius of the cavity, $r^2 = \xi^2 + y^2 + z^2$, and the constant of integration is chosen so that $\Phi(R) = 1$.

The electromagnetic fields inside the relativistic cavity are

$$E_x = \xi/2, \quad E_y = -B_z = y/4,$$
$$B_x = 0 \quad E_z = B_y = z/4. \tag{14}$$

The calculated distribution of electromagnetic fields is close to the one observed in the 3D PIC simulation (see Fig. 2). The small deviation from the analytically calculated field distribution is because the cavity shape is not exactly a sphere. It is easy to see that the fields (14) satisfy the Maxwell equations.

The Lorentz force acting on a relativistic electron with $v_x = 1$ inside the cavity is

$$F_x = -\frac{\partial \Phi}{\partial \xi} = -E_x = -\frac{\xi}{2}, \tag{15}$$

$$F_y = -\frac{\partial \Phi}{\partial y} = -E_y + B_z = -\frac{y}{2}, \tag{16}$$

$$F_z = -\frac{\partial \Phi}{\partial z} = -E_z - B_y = -\frac{z}{2}. \tag{17}$$

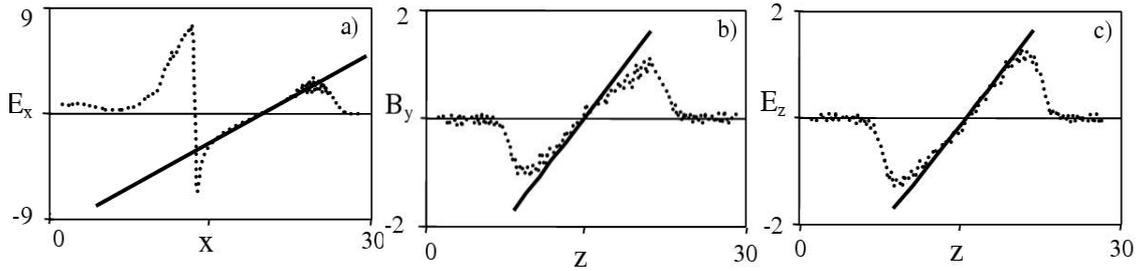

**Figure 2.** Space distribution of the electromagnetic fields normalized to $mc\omega_p/|e|$ at the time instance when the laser pulse has passed $25c/\omega_p$: (a) $E_x$ as a function of $x$; (b) $B_y$ as a function of $z$; (c) $E_z$ as a function of $z$. The PIC simulation results are shown by dashed lines while the analytical results are shown by solid lines. The coordinates are given in $c/\omega_p$.

The wake potential, $\Phi$, can be considered as the potential of the Lorentz force on the electron with $v_x = 1$. The Lorentz force peaks for the electron with $v_x = v_0 = 1$ while it is zero for the electron with $v_x = -1$ because of the relativistic compensation of the electrostatic force by the self-magnetic force. Notice that this effect is opposite to that of the relativistically moving ionic sphere. This is because the displacement current in the cavity is opposite to the ion current in the relativistically moving ion sphere.

## 4. SCALABILITY OF THE BUBBLE

One of the most intriguing questions is the parameter region, where the bubble exists and how the electron energy spectrum depends on the laser intensity and the plasma density. Although there is a quite developed analytical theory for the weakly relativistic regime, $a \ll 1$, very little is known in the ultra-relativistic regime, $a \gg 1$. Yet, it is the ultra-relativistic laser-plasma regime, where the bubble is formed. The analytical difficulty here is that the electrons form multi-stream flows, which must be described in the fully kinetic way.

Here we show that in spite of the physics complexity, a very general and analytically accurate scaling of the bubble can be derived. For this purpose we have to solve the Vlasov equation on the electron distribution function $f(t, \mathbf{r}, \mathbf{p})$

$$\left[\partial_t + \mathbf{v}\partial_\mathbf{r} - e\left(\mathbf{E} + \frac{\mathbf{v}}{c} \times \mathbf{H}\right)\partial_\mathbf{p}\right] f(t, \mathbf{p}, \mathbf{r}) = 0, \tag{18}$$

together with the Maxwell equations on the electric **E** and magnetic **H** fields. A simple dimension analysis yields

$$f = \frac{n_e}{(mc)^3} F\left(\omega_0 t, \frac{\mathbf{p}}{m_e c}, \frac{\omega_0 \mathbf{r}}{c}, \frac{n_c}{n_e}, a_0, \frac{R}{\lambda}, \omega_0 \tau\right), \tag{19}$$

where $F$ is an unknown universal function and $R$ is the radius and $\tau$ is the duration of the laser beam. Eq. (19) is of little use as long as it depends on four dimensionless parameters. However, we are interested only in the ultrarelativistic limit. Thus, we can set $\mathbf{v} = c\mathbf{p}/|\mathbf{p}|$ and re-write the Vlasov equation as

$$\left[\partial_t + \frac{\mathbf{p}}{|\mathbf{p}|}\partial_{\mathbf{r}} - q\left(\mathbf{E} + \frac{\mathbf{p}}{|\mathbf{p}|} \times \mathbf{H}\right)\partial_{\mathbf{p}}\right] f = 0. \tag{20}$$

Further, we make the variables dimensionless in the standard way: $t \to \omega_0 t$, $\mathbf{p} \to \mathbf{p}/mca_0$, $\mathbf{r} \to \omega_0 \mathbf{r}/c$, $(\mathbf{E}, \mathbf{H}) \to (\mathbf{E}, \mathbf{H})/\omega A_0$ and re-write Eq. (20) together with the Maxwell equations in the dimensionless form

$$\left[\partial_t + \frac{\mathbf{p}}{|\mathbf{p}|}\partial_{\mathbf{r}} - \left(\mathbf{E} + \frac{\mathbf{p}}{|\mathbf{p}|} \times \mathbf{H}\right)\partial_{\mathbf{p}}\right] f_S = 0;$$
$$\nabla_{\mathbf{r}} \cdot \mathbf{E} = S(1 - \rho), \qquad \nabla_{\mathbf{r}} \cdot \mathbf{H} = 0, \tag{21}$$
$$\nabla_{\mathbf{r}} \times \mathbf{H} = S\mathbf{j} + \partial_t \mathbf{E}, \qquad \nabla_{\mathbf{r}} \times \mathbf{E} = -\partial_t \mathbf{H},$$

where

$$\rho = \int f_S \, d\mathbf{p}, \quad \mathbf{j} = \int (\mathbf{p}/|\mathbf{p}|) f_S \, d\mathbf{p}. \tag{22}$$

Eqs. (21) contain the only one dimensionless parameter $S = n_e/a_0 n_c$ and the unknown universal function $f_S(t, \mathbf{p}, \mathbf{r}) = (m_e^3 c^3 a_0^3/n_e) f(t, \mathbf{p}, \mathbf{r})$ Now we can express the electron distribution function:

$$f = \frac{n_e}{(m_e c a_0)^3} f_S(t, \mathbf{p}, \mathbf{r}, R, \tau, S). \tag{23}$$

The universal function in Eq. (23) has no explicit dependence on $n$ and $a$, but rather depends on the dimensionless similarity parameter $S$ only.

Thus, if we fix the laser radius $R$ and duration $\tau$, and change simultaneously $a_0$ and $n_e$ so that $S = const$, then the dynamics of the ultra-relativistic laser plasma scales identically. This means that if we have found a bubble at some particular values of laser pulse energy $W_0$, amplitude $a_0$ and $n_0$, then we can scale it to larger or lower laser intensities if we keep $S = n/an_c = const$. In this scaling, the laser energy $W$ changes as

$$W = \left(\frac{a}{a_0}\right)^2 W_0, \tag{24}$$

the electron relativistic $\gamma$−factor scales as

$$\gamma = \frac{a}{a_0}\gamma_0, \tag{25}$$

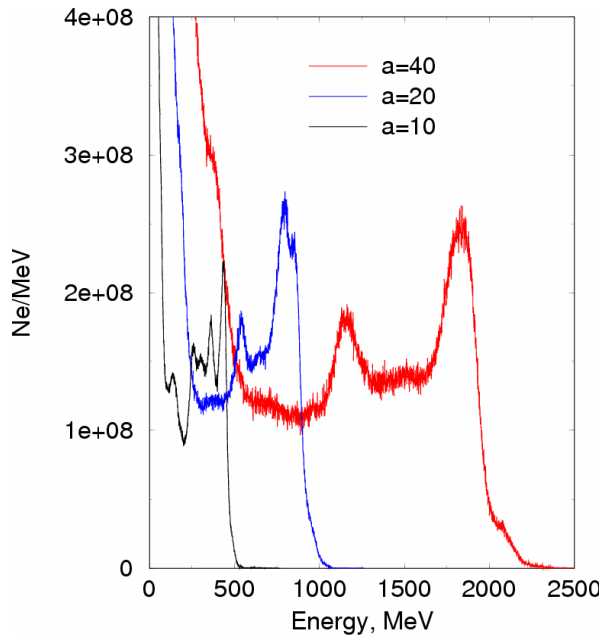

**Figure 3.** Upscaling the bubble. Electron energy spectra obtained for lasers with $a = 10, 20, 40$ are shown. The similarity parameter $S = n/an_c$ has been kept constant. The spectra are similar and can be transformed from one into another by stretching the energy axis proportionally to the amplitude $a$.

the number of accelerated electrons $N$ scales as

$$N = \frac{a}{a_0} N_0. \tag{26}$$

We have compared this analytical scaling with direct 3D PIC simulations using the code VLPL [15]. As the generating bubble we took a 30 fs laser pulse with $W_0 = 10$ J energy focused to the amplitude $a_0 = 10$ propagating in plasma with $n_0 = 0.01 n_c$ over $1000\lambda$ distance [14]. To upscale the bubble, we repeated the simulation with laser pulses having the same shape, but with the amplitudes $a_1 = 20$ (energy $W_1 = 40$ J) and $a_2 = 40$ (energy $W_2 = 160$ J) running in plasmas with $n_1 = 0.02 n_c$ and $n_2 = 0.04 n_c$. The electron energy spectra obtained after $1000\lambda$ propagation distance are shown in Fig. 3. The obtained spectra satisfy the similarity (24)-(26). To transform one spectrum into another, one may simply stretch them along the energy axis proportionally to the laser amplitude $a$. At the same time, the number of energetic electrons scales as $a$.

Now, as soon as we are able to upscale the bubble, we may try to downscale it as well. Although the downscaling would mean entering a moderately relativistic regime, where the similarity theory is less reliable. The LOA experiment [21] has been done with 1 J energy, 30 fs laser pulse. According to the similarity theory (24)-(26) the plasma density should be $n \approx 0.003 n_c$ and the monoenergetic peak should be expected at $\approx 170$ MeV. The spectrum obtained in a 3D VLPL simulation for the LOA parameters is shown in Fig. 4. We see that the electron energy is well scaled, but the number of electrons is somewhat less than that expected from the similarity theory. Yet, in this moderately relativistic regime even this scalability can be valuated as surprisingly good.

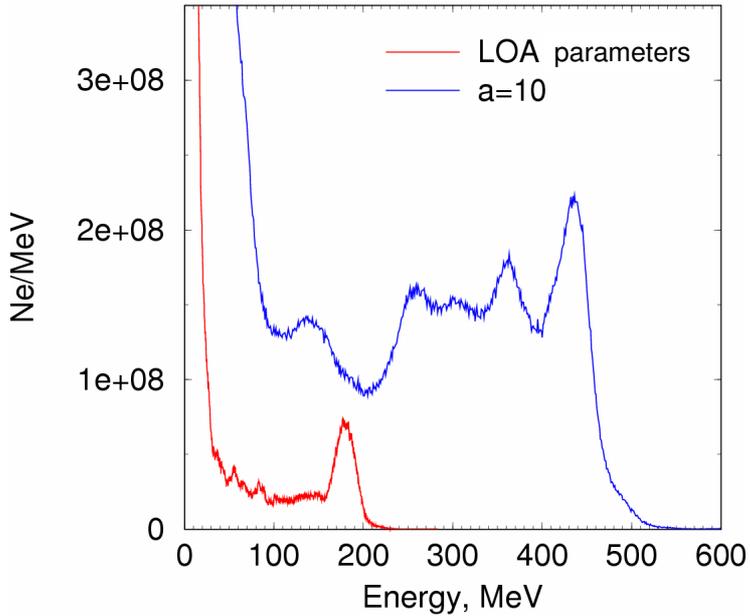

**Figure 4.** Downscaling the bubble. The similarity parameter $S = n/an_c$ has been kept constant and the laser energy has been downscaled to 1 J (LOA laser parameters).

## 5. CONCLUSIONS

The ultra-relativistic laser-plasma bubble is a very promising regime of acceleration, because it leads to quasi-monoenergetic electron beams. In the present work we derive an analytical expression for the fields within a spherical cavity moving at relativistic velocity in plasma. We show that the fields linearly depend on the coordiantes and act to compress transversely the co-propagating electron beam.

We develop a similarity theory and show that the ultra-relativistic plasma has a non-trivial similarity parameter $S = n/an_c$. According to this parameter we check the bubble scalability numerically.

One of the authors (I. K.) gratefully acknowledges the hospitality of the Institute for Theoretical Physics of Duesseldorf University. This work has been supported in parts by the Alexander von Humboldt Foundation, DFG and BMBF (Germany), and by the Russian Fund for Fundamental Research (Grants No 01-02-16575, No 01-02-06488 and by Russian Academy of Science (Grant N 1999-37).